\documentclass[onecolumn,showpacs,showkywords,preprintnumbers,amsmath,amsfonts,mathrsfs,amssymb,preprint]{revtex4}

\usepackage[latin1]{inputenc}
\usepackage[pdftex]{graphicx}
\usepackage{epsfig}
\usepackage{graphicx}
\usepackage{dcolumn}
\usepackage{bm}
\usepackage{color}
\usepackage{natbib}
\newcommand {\be}{\begin{equation}}
\newcommand {\ee}{\end{equation}}
\newcommand {\ba}{\begin{eqnarray}}
\newcommand {\ea}{\end{eqnarray}}

\usepackage{amsmath}  
\usepackage{amsfonts} 
\usepackage{graphicx} 
\usepackage{nicefrac}
\usepackage{hyperref}

\begin{document}


\title{Carnot, Stirling, Ericsson stochastic heat engines: Efficiency at maximum power} 

\author{O. Contreras-Vergara}
\author{N. S\'anchez-Salas}
\affiliation{Departamento de F\'isica,
Escuela Superior de F\'isica y Matem\'aticas, Instituto Polit\'ecnico Nacional, 
 Edif. 9 UP Zacatenco, CP 07738, CDMX, M\'exico.}
 \author{G. Valencia-Ortega}
 \affiliation{Departamento de Biof\'isica,
Escuela Nacional de Ciencias Biol\'ogicas, Instituto Polit\'ecnico Nacional, 
UP Santo Tom\'as, C.P. 11340, CDMX, M\'exico.}
\author{J. I. Jim\'enez-Aquino}
\affiliation{Departamento de F\'{\i}sica, Universidad Aut\'onoma Metropolitana--Iztapalapa, 
C.P. 09340, CDMX, M\'exico.}


\begin{abstract}
This work obtains the efficiency at maximum power for a stochastic heat engine performing Carnot-like, Stirling-like and Ericsson-like cycles. For the mesoscopic engine a Brownian particle trapped by an optical tweezers is considered. 
The dynamics of this stochastic engine is described as an overdamped Langevin equation with a harmonic potential, whereas is in contact with two thermal baths at different temperatures, namely, hot ($T_h$) and cold ($T_c$). The harmonic oscillator Langevin equation is transformed into a macroscopic equation associated with the mean value $\langle x^2(t)\rangle$ using the original Langevin approach. At equilibrium stationary state this quantity satisfies a state-like equation from which the thermodynamic properties are calculated. To obtained the efficiency at maximum power it is considered the finite-time cycle processes under the framework of low dissipation approach.

\end{abstract}
\maketitle 

\section{Introduction}

With the use of nano-technology, physicists and engineers attempt to overcome the challenge of building artificial mesoscopic machines capable of extracting energy from their environment, and convert it into useful work. 
The main characteristic of these micro- or nano engines is that the magnitude of the fluctuations generated by their environment is comparable to the average flow of energy produced by such engines. Then, the performance of a mesoscopic engine strongly depends on the properties of its surroundings. For instance, the so-called Brownian Motors \cite{Reimann1996, Reimann2002, Hanggi2005} are  devices capable of rectify fluctuations to produce useful work; they have a ratchet-like design in which a spatial anisotropic or asymmetric potential is involved, and an additional ingredient that takes the system out of equilibrium. The ratchet model proposed by Feynman has been used as an inspiration for the study of a significant number of theoretical and experimental works on Brownian motors and other devices \cite{Magnasco1993,Reimann1996, Julicher1997,  Qian1997, Bier1997, Reimann2002, Hanggi2005, Lacoste2007, Perez2010, Goychuk2014, Tu2018, Hwang2019, Caballero2020, Gulyaev2020} like the stochastic heat engines in the construction of microscopic devices. 
The first experimental realization of a microscopic heat engine, comprising a single colloidal particle subject to a time-dependent optical trap was reported in \cite{Blickle12}. 
In this experiment, the  Brownian particle performs a Stirling-like cycle, in  which the particle and the trapping potential replace the working gas and the piston of its macroscopic counterpart.  
The technique of trapping particles using optical tweezers has been shown to provide a precise control over the confinement and  temperature of the colloidal particle \cite{Gieseler2014, Martinez2015}.  

The stochastic heat engines  can operate cycles between isothermal baths; however, recent experiments have been made  with colloidal  engines that operate in cycles between different thermal baths including non-isothermal processes, both in active and passive media \cite{ Blickle12, Martinez2015,Martinez16, DeLorenzo15, Thorneywork20}.  The two fundamental quantities used to characterize the thermodynamic properties of macroscopic and microscopic heat engines are both the power and efficiency. The main difference between both devices is that in the latter the stochastic fluctuations play a fundamental role. 
At these days, stochastic thermodynamics has been the appropriate theoretical framework  to characterize the energetics of microscopic heat engines in which the concepts of work, heat, internal energy, etc., have been defined along a single stochastic trajectory \cite{Sekimoto10}. 

It is well known that for macroscopic heat engines the Carnot efficiency can be achieved when the cyclic process is quasistatic, which also means a large cycle time  and zero power output. The so-called finite-time thermodynamics is related to the study of  the efficiency of heat engines when the cyclic process is performed at finite-time. It seems that the pioneering work related to the efficiency at maximum power of an endoreversible engine operating at finite-time was reported by Novikov \cite{Novikov1958}. However, the work by Curzon and Ahlborn \cite{Curzon1975} is more cited in the literature in which 
the efficiency at maxium power is shown to satisfy  $\eta_{_{CA}}=1-\sqrt{T_c/T_h}$, which is less than the Carnot efficiency $\eta_{_{CA}} < \eta_{_C}$, for an engine operating between two heat baths at temperatures $T_h$ and $T_c$ ($T_c<T_h$). Since then, a significant number of papers related to Finite-Time Thermodynamics have been reported in the literature \cite{Hoffmann1997,Chen2004,Andresen2011,Bejan2013}, and all references there in. 

In this context, it is worth to comment the paper by Esposito et al. \cite{Esposito10}, related to the efficiency at maximum power for macroscopic heat engines performing finite-time Carnot cycles and operating under low dissipation conditions. According to the authors, 
the starting point of low-dissipation approach is a Carnot engine which operates under reversible conditions when the system always remains close to equilibrium and the cycle time become very large.  If the cycle processes are no longer reversible but irreversible at finite-time, then the dissipative processes play an important role being the low-dissipation limit an interesting theoretical approach to characterize them. This approach establishes that, if $\tau_c$ ($\tau_h$) are the time in which the system is in contact with the cold (hot) reservoir along a cycle, then the entropy production per cycle along the cold (hot) part of the cycle is proposed to behave as $\Sigma_c/\tau_c$ ($\Sigma_h/\tau_h$), where $\Sigma$ 
contains information about the dissipation or irreversibility present at isothermal branches.  Therefore, the amount of heat per cycle incoming to the system from the cold (hot) reservoir are given by $Q_c=T_c(-\Delta S-\Sigma_c/\tau_c + \cdots)$ and $Q_h=T_h(\Delta S-\Sigma_h/\tau_h+\cdots)$, being $Q^c_{\infty}=-T_c\Delta S$ and $Q^h_{\infty}=T_h\Delta S$, the amount of heat exchanged with the cold (hot) reservoir under reversible conditions. Whereas that the adiabatic processes are considered as instantaneous and therefore the irreversible effects are only taken into account in the two isothermal processes.

The study of the efficiency at maximum power has also been extended to stochastic heat engines, as can be corroborated in the recent literature 
\cite{ Schmiedl07,Tu14,Plata19,Blickle12,Viktor2015}.
Due to the aforementioned features of a stochastic heat engine, the  theoretical model used for its description is the Langevin equation in which the optical trap is represented by a harmonic potential. The model also considers the engine (the particle) in contact with two thermal baths at different temperatures $T_h$ (hot) and $T_c$ (cold), with $T_h>T_c$, and the internal noise intensity changes in time from the hot to cold  values. The stochastic efficiency is also defined as the ratio of the stochastic work extracted in a cycle and the stochastic heat transferred from the hot bath to the particle in a cycle  \cite{Martinez2015}. \\
In a work published by Blickle \cite{Blickle12}, 
it was shown that at thermodynamic level the mean efficiency $\langle \eta(\tau)\rangle$ is less  than the Carnot efficiency $\eta_{_C}=1-T_c/T_h$, that is $\langle \eta(\tau)\rangle=\langle W(\tau)\rangle/\langle Q(\tau)\rangle <\eta_{_C}$, where $\tau$  is the cycle time, $\langle W(\tau)\rangle$ the average work and $\langle Q(\tau)\rangle$ the average heat flux.
Also, the  experiments \cite{Blickle12,Martinez2015} show that for shorter cycle times, the dissipation effects become important, and the mean work per cycle can be written as $\langle W\rangle= \langle W_{\infty}\rangle +\langle W_{dis}\rangle$, where $\langle W_{\infty}\rangle$ is the mean quasistatic work for longer cycle times and $\langle W_{dis}\rangle$ the mean irreversibly dissipated work per cycle. This latter can be written to first order as $\langle W_{dis}\rangle=\Sigma/\tau$ where the coefficient  $\Sigma$, contains information about the irreversibilities present in the cycle, for instance, time-dependent protocol and the coupling mechanism between the particle and the thermal environment. In other words,  $\Sigma/\tau$ accounts for an amount of energy dissipated in a cycle. 
In \cite{Blickle12} it has also been commented that, in the experiment at small scales it is very difficult to keep  hot and cold reservoirs thermally isolated, so rather than coupling the colloidal
particle periodically to different heat baths, the temperature of the surrounding liquid is suddenly
changed.

The purpose of the present contribution is to apply the 
low-dissipation considerations to obtain the efficiency at maximum power of a Brownian heat engine, which can operate in three finite-time irreversible cycles between a hot and a cold reservoir at temperatures $T_h$ and $T_c$, respectively. Three different cycles namely, Carnot-, Stirling-, and Ericsson-like are considered. 
The theoretical analysis is formulated in the context of a Langevin approach for a Brownian particle in a harmonic trap with time-dependent stiffness $\kappa(t)$. The strategy is as follows: first, the Langevin equation is transformed into a macroscopic one for the average value $\langle x^2(t)\rangle$, in  which the time-dependent temperature $T(t)$ is also taken into account (both the stiffness and temperature are externally controlled \cite{Plata20}). Instead of solving the macroscopic deterministic equation for specific protocols $\kappa(t)$, and $T(t)$, the advantage from the system equilibrium thermodynamic properties is taken into account by means of the \textit{state-like equation} associated with the average value $\langle x^2\rangle_{eq}$. This allows to obtain the work, heat and efficiency under quasistatic  conditions, and the irreversible effects, coming from both the control of the stiffness potential and bath temperature, can be taken into account through the dissipation parameter $\Sigma$ \cite{Blickle12}. Once this is done, the efficiency at maximum power characterized by finite-time cycles can be obtained using the low-dissipation approach. 

This work is organized as follows: Section II obtains the macroscopic equation  (ensemble property) for the overdamped harmonic oscillator associated with the mean value $\langle x^2(t)\rangle$, using the original method proposed by Langevin in 1908 \cite{Langevin1908,Lemons1997, Contreras2021}. Then\textcolor{red}{,} by means of the state-like equation,  the equilibrium thermodynamic properties for the Carnot-, Stirling-, and Ericsson-like cycles are calculated. Section III focuses on the study of low-dissipation approach to calculate the efficiency at maximum power of each heat engine, and the theoretical results are compared with other reported results. Our conclusions and comments are given in Section IV.  \\

\section{Stochastic heat engine} 

The mathematical model proposed to describe the dynamics of a Brownian heat engine in contact with a thermal bath is given by a Langevin equation for a Brownian harmonic oscillator with time-dependent stiffness $\kappa(t)$, that is
\be
m\frac{d^2 x}{dt^2}=-\alpha \frac{dx}{dt} - \kappa(t) x + \xi(t)  , \label{hole}
\ee
where $m$ is the particle mass, $\alpha=6\pi \zeta a$ the friction coefficient, being $\zeta$ the fluid viscosity and $a$ the radius of the particle assumed to be a sphere. In the overdamped regime it becomes    
\be 
\alpha\frac{dx}{dt}=-\kappa(t) x+ \xi(t) .   \label{odho} \ee
This stochastic differential equation rules a process well  known as Ornstein-Uhlenbeck  and it becomes stationary in long time limit, $t\to\infty$, for which $\kappa(t)\to \kappa=$const. This stochastic differential equation can be transformed into a macroscopic one following Langevin's strategy as reported in \cite{Contreras2021}. 
From Eq. (\ref{odho}) it is straightforward to obtain the equation
\begin{equation}
    \alpha\frac{d\langle x^2\rangle }{dt}=-2\kappa(t)\langle x^2 \rangle + {2} \langle x\xi(t)\rangle . \label{msd1}
\end{equation}

This equation becomes:  
\begin{equation}
    \alpha  \frac{d \langle x^2\rangle}{dt}=-2\kappa(t) \langle x^2\rangle + 2k_{_B}T. \label{dex2}
\end{equation}
where $\langle x\xi(t)\rangle=k_{_B}T$, being $k_{_B}T$ the intensity of thermal noise and $T$ the equilibrium bath temperature.  However, we can also suppose that the correlation function is not a constant but a time-dependent function through the temperature, that is, $\langle x\xi(t)\rangle=k_{_B}T(t)$, and thus the macroscopic Langevin equation can be written as
\begin{equation}
    \alpha  \frac{d \langle x^2\rangle}{dt}=-2\kappa(t) \langle x^2\rangle + 2k_{_B}T(t). \label{dex2b}
\end{equation}
It is clear that in the equilibrium stationary state $\langle x^2\rangle_{st}=k_{_B}T/\kappa$, because $T(t)\to T$ as the time gets large. In several and recent papers \cite{Schmiedl08,Tu2018, Rana2014,Plata19,Viktor2015}, the amount of work, the interchange heat with the thermal bath as well as the efficiency at maximum power performed by a Brownian heat engine during a cycle are calculated along a single stochastic trajectory taking into account a specific form of the time-dependent protocol $\kappa(t)$.

In the present work, instead of solving Eq. (\ref{dex2b}) for specific protocols for  $\kappa(t)$ and $T(t)$,  a different strategy related to low-dissipation approach to obtain the efficiency at maximum power for three stochastic heat engines. In our case, the irreversible average work and irreversible average heat will be given respectively by $\langle W\rangle \approx \langle W\rangle_{\infty}+ \frac{\Sigma}{\tau}$ and  $\langle Q\rangle\approx \langle Q\rangle_{\infty}-\frac{T\Sigma}{\tau}$, where $\langle W\rangle_{\infty}$ and 
$\langle Q\rangle_{\infty}$ are the average work and average heat, respectively,  under equilibrium conditions. Here the parameter $\Sigma$ takes into account all information coming from any of irreversibility sources including the time-dependent protocols.

\subsection{Quasistatic description of Brownian heat engines}
 
 The thermodynamics properties of the Brownian heat engine, can be calculated by means of the {\it equation of state} in a similar way as in the case of an ideal gas in classical thermodynamics. A state-point is characterized by $(\langle x^2\rangle, \kappa, T)$ as thermodynamic variables, and the stiffness of the optical trap as well as the bath temperature can be considered as time-independent quantities (in what follows we consider $\langle x^2\rangle_{st}\equiv\langle x^2\rangle$). Likewise, the average of the total energy $\langle E\rangle\equiv E$, is proposed to satisfy the equation  \cite{Plata19}
\begin{equation}
E={1\over 2} \kappa \langle x^2\rangle + {1\over 2}k_{_B}T ,  \label{ae}   \end{equation}
and therefore $E_{eq}=k_{_B}T$.  
Once defined the energy available by the system, different thermodynamic-like processes can be explored whereas the Brownian particle is in contact with to thermal baths at different temperatures, hot $T_h$ and cold $T_c$  \cite{Martinez16,Blickle12,Quinto14,Krishnamurthy16}. To calculate the efficiency at quasistatic conditions for the three aforementioned stochastic heat engines, we proceed as follows. According to Sekimoto \cite{Sekimoto10}, the thermodynamic first law-like along a stochastic trajectory reads, in the overdamped regime, as $dE=d^{\prime}Q+d^{\prime}W$, where $dE=dU$, being $U$ the potential energy and $d^{\prime}W={\partial U\over\partial \lambda}d\lambda$, with $\lambda$ an external parameter. In our case $\lambda=\kappa$,  $U(x,\kappa)={1\over 2}\kappa\, x^2$ and $d^{\prime}W={1\over 2}x^2 d\kappa$. However, the average work as well the average heat are the same as the thermodynamic quantities, that is,  $W=\langle W\rangle$ and $Q=\langle Q\rangle$,  and according to Eq. (\ref{ae}) it can be shown that $d^{\prime}Q={1\over 2}\kappa d\langle x^2\rangle + {1\over 2}k_{_B} dT$. 

The total work $W$ and heat $Q$ exchanged with the surroundings along a quasistatic trajectory, from a one state A to another state B, are given by, respectively
\ba
W_{AB} &=& {1\over 2}\int _A^B \langle x^2\rangle \, d\kappa ,  \label{Wa}   
\\
Q_{AB}&=&{1\over 2}\int_{A}^{B}  \kappa d\langle x^2\rangle + {1\over 2}k_{_B} (T_B -T_A).  \label{Qa}
 \ea
The Brownian particle free energy can be obtained from the  partition function given by  $ Z(\kappa,T)= \sqrt{ 2 \pi k_{_B}T/\kappa}$. The free energy $F(\kappa,T)$ thus becomes 
$F(\kappa,T)=-k_{_B}T \ln\sqrt{ 2 \pi k_{_B}T / \kappa}$. Likewise, in analogy to the differential form of the thermodynamic potential for a ideal gas, $dF=-SdT-pdV$ there is a correspondence with ensembles of a single-confined colloidal particle in the form: $dF=-SdT+\Phi d\kappa$, where the entropy $S$ reads 
$S=-\left({\partial F\over\partial T}\right)_\kappa 
=(k_{_B}/2)[\ln(2 \pi k_{_B}T/ \kappa)+1]$, and the auxiliary conjugate thermodynamic variable  $\Phi$, also satisfies the state-like equation   
$\Phi= \left( {\partial F\over\partial \kappa}\right)_T={k_{_B}T\over 2\kappa} ={\langle x^2\rangle \over 2}$. 
In this equation the trap stiffness is analogous to the inverse of the effective volume in the state equation of an ideal gas $p\sim T/V$,
while the intensive variable $\Phi$, related to the variance of the particle trajectories can be seen as a kind of effective macroscopic pressure. We complement our theoretical study with the calculation of the adiabatic-like equation associated with the system, which is easily obtained from the condition $d^{\prime}Q=0$, yielding to $\kappa={\rm const} \, T^2$ consistent with what was previously reported in \cite{Sekimoto00,Martinez2015}.
These guidelines specify the thermodynamic processes analogous to those established in macroscopic systems. Two thermodynamic variables are usually considered to characterize the performance of those types of robust heat engines, the extracted work $W$ and the conversion efficiency $\eta$. 

In analogy with real-macroscopic heat engines, the energy conversion efficiency of thermodynamic protocols with single particles in suspension is constrained by the second law of thermodynamics \cite{Sekimoto00,Schmiedl07,Gingrich16}, that is,
\begin{equation}
    \eta \equiv \frac{\langle W(\tau) \rangle}{\langle Q_h(\tau) \rangle} \leq \eta_{_C}, \label{effreffc}
\end{equation}
where $\eta_{_C}$ is the Carnot efficiency. In thermal equilibrium conditions, i.e, in the quasistatic limit ($\tau \to \infty$), the efficiency for these types of block-thermodynamic cycle models are closed to $\eta_C$, under this condition the power output is zero.

Isothermal paths can be defined as thermodynamic processes where the temperature of the surroundings remains constant in time. In optically trapped particle systems immersed in a fluid,  isothermal processes are associated with the so-called breathing optical parabola \cite{Martinez13,Dinis16}. That is, as the stiffness of the trap is decreased (increased) space for the particle increases (decreases).  Thus, the free energy change $\Delta F$ stands for the useful reversible work that thermodynamic protocols can performed during the expansion and compression processes at fixed temperatures.

\subsection{Carnot-like cycle}

\begin{figure}
    \centering
    \includegraphics[scale=1.1]{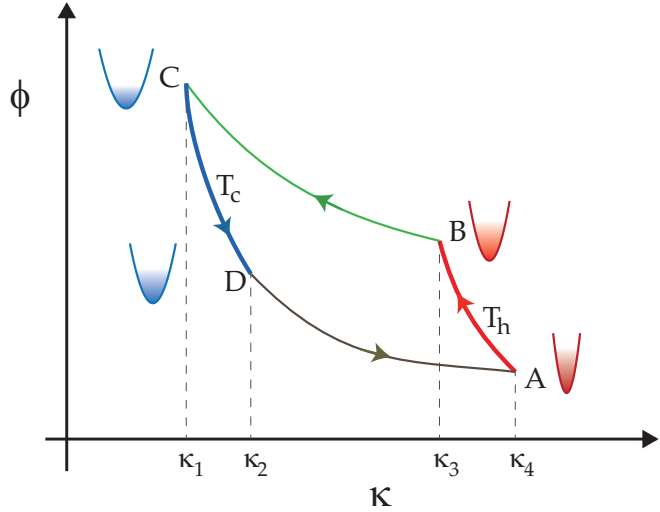}
    \caption{$\Phi$-$\kappa$ thermodynamic diagram of a Carnot-like cycle where (i) an isothermal expansion (red path); (ii) an adiabatic expansion (B-C green path); (iii) an isothermal compression (blue path) and (iv) an adiabatic compression (D-A green path). }
    \label{Carnot}
\end{figure}

The Carnot Cycle represents the paradigm of thermal cycles, it consists of two adiabatic and two isothermal branches, whose maximum efficiency is, $\eta_C=1-T_c/T_h$, under reversible conditions. The Carnot-like cycle for a Brownian particle can be implemented modifying the stiffness $\kappa$ and the bath temperature $T$ \cite{Martinez2015}, see Fig. \ref{Carnot}. The idealized Carnot cycle involves two quasistatic isothermal processes where $T$ is kept constant but $\kappa$ also changes  quasistatically, and two reversible adiabatic processes where $T$ and $\kappa$ change, but along the adiabatic path, $\kappa ={\rm const}* T^2$. The energetic description of this Carnot-like cycle can be summarized as follows:

i) Isothermal expansion process ($A\to B$): In this process the stiffness potential trap decreases from $\kappa_4\to \kappa_3$ ($\kappa_3<\kappa_4$) at $T_h={\rm const}$. Besides, $(\Delta E)_{AB}=0$ and $Q_{AB}=-W_{AB}$,
\ba
W_{AB} &=& {k_{_B}T_h\over 2}\int _{\kappa_4}^{\kappa_3} {d\kappa\over \kappa} = 
 {k_{_B}T_h\over 2}\ln\left(\kappa_3\over \kappa_4\right)<0 ,   \label{wab} \\
 \nonumber\\
 Q_{AB}&=& -{k_{_B}T_h\over 2}\ln\left(\kappa_3\over \kappa_4\right) ={k_{_B}T_h\over 2}\ln\left(\kappa_4\over \kappa_3\right) >0 . \label{cqab}
 \ea

ii) Adiabatic expansion process ($B\to C$): An amount of work is performed from $\kappa_3\to \kappa_1$, ($\kappa_1 <\kappa_3$), while $T$ change from $T_h \to T_c$. As $Q_{BC}=0$ then
$W_{BC}=(\Delta E)_{BC}=k_{_B}(T_h-T_c)$ 

iii) Isothermal compression process ($C\to D$): This process takes place now for increasing values of $\kappa$ from 
$\kappa_1\to \kappa_2$ ($\kappa_2 > \kappa_1$),
while the system is in contact with thermal bath at $T=T_c$, where $Q_{CD}=-W_{CD}$; therefore,
\ba
W_{CD} &=& {k_{_B}T_c\over 2}\int _{\kappa_1}^{\kappa_2} {d\kappa\over \kappa} =
 {k_{_B}T_c\over 2}\ln\left(\kappa_2\over \kappa_1\right)>0 ,   \label{cwcd} \\
 \nonumber\\
 Q_{CD}&=& -{k_{_B}T_c\over 2}\ln\left(\kappa_2\over \kappa_1\right) <0 . \label{cqab1}
 \ea

iv) Adiabatic compression process ($D \to A$): The last amount of work is extracted from $\kappa_2\to \kappa_4$ (with $\kappa_4 > \kappa_2$), in which $Q_{DA}=0$, and thus 
$W_{DA}=(\Delta E)_{DA}=k_{_B}(T_h-T_c)$.

The total work performed by this block-thermodynamic cycle becomes $W_{tot} = W_{AB}+W_{BC}+W_{CD}+W_{DA}$, and according to Eq. (\ref{effreffc}) the efficiency
reads 
\be
\eta={W_{tot}\over Q_{in}}= \frac{ {k_{_B}\over 2}\left[T_h\ln\left({\kappa_4\over \kappa_3}\right) - T_c\ln\left({\kappa_2\over \kappa_1}\right)\right]}{{k_{_B} \over 2} T_h \ln\left( {\kappa_4\over \kappa_3}\right)}, \label{cea}
\ee
being $Q_{in}=Q_{AB}$. However, from  adiabatic equation and according to Fig. (\ref{Carnot}), its 
\be
  {\kappa_3\over T_h^2}={\kappa_1 \over T_c^2}  \qquad {\kappa_4 \over T_h^2}={\kappa_2\over T_c^2} \qquad \text{thus} \qquad {\kappa_3\over \kappa_1}={\kappa_4\over \kappa_2},\label{ks}  
\ee
and therefore the Carnot-like efficiency becomes  
\be
\eta_{_C}={T_h-T_c\over T_h } =1-{T_c\over T_h} ,    \label{ks1}  
\ee
which is an expected result. As in the thermodynamic case, the efficiency only depends on the temperatures of the thermal baths. 

\subsection{Stirling-like Cycle}

\begin{figure}
    \centering
    \includegraphics[scale=1.15]{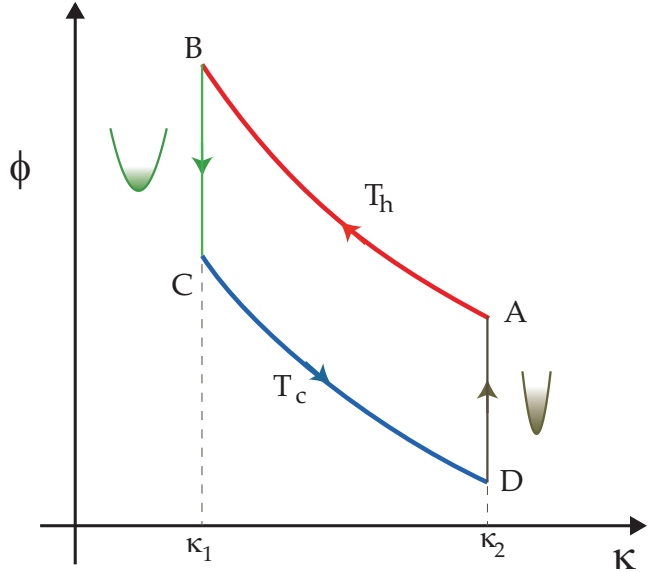}
    \caption{$\Phi$-$\kappa$ thermodynamic diagram of a Stirling-like cyclewhere (i) an isothermal expansion (red path); (ii) an isocoric cooling process B-C (green path) ; (iii) an isothermal compression (blue path) and (iv) an isocoric heating process (D-A green path). }
    \label{Stirling}
\end{figure}

In this subsection the analysis of a Stirling-type cycle is presented. The first reported experimental micro-size heat engine was built inspired in this cycle \cite{Blickle12}. Analogously to the macroscopic case, this cycle is composed of two isothermal  linked through two isochoric processes  (two pair of symmetric processes at $\kappa={\rm const.}$, as shown  in Fig. \ref{Stirling}). Each path is swept in a quasistatic way going between two equilibrium states. For a Brownian particle in a harmonic trap, the process is carried out under certain conditions of stiffness and temperature. The energetic in each trajectory can be stated as follows:\\

i) Isothermal expansion process ($A\to B$): 
During this process the stiffness of the trap changes from $\kappa_2$ to $\kappa_1$ ($\kappa_1 < \kappa_2$) and $T_h={\rm const}$. Besides, $\Delta E_{AB}=0$ and $Q_{AB}=-W_{AB}$, with
\begin{equation}
    W_{AB} = {k_{_B}T_h\over 2} \ln \left ( \kappa_1 \over \kappa_2 \right )
\end{equation}

ii) Isochoric process ($B \to C$): In this case the stiffness remains constant ($\kappa_1={\rm const.}$) and the potential does not change; the work vanishes $W_{BC}=0$, and the heat is $Q_{BC}=(\Delta E)_{BC}$, such that
\be
Q_{BC}=k_{_B}(T_c-T_h)<0   . \label{qbc}    \ee

iii) Isothermal compression process ($C\to D$): During this process the stiffness of the trap changes from $\kappa_1$ to $\kappa_2$  and $T_c={\rm const.}$, the colder temperature. In this case,  $(\Delta E)_{CD}=0$ and $Q_{CD}=-W_{CD}$, where 
\be
    W_{CD} = {k_{_B}T_c\over 2} \ln \left ( \kappa_2 \over \kappa_1 \right ). 
\ee

iv) Isochoric process ($D\to A$): Similarly to the second branch, the stiffness remains constant ($\kappa_2={\rm const}$) and the potential does not change; the work vanishes $W_{DA}=0$, and the heat is $Q_{DA}=(\Delta E)_{DA}$, such that 
\be
Q_{DA}=k_{_B}(T_h-T_c)>0    .\label{qda}
\ee
The efficiency of this cycle is given by
\ba
\eta &=& \frac{W_{tot}}{Q_{in}} ,
 \nonumber\\
 &=& \frac{{k_{_B}\over 2}\left[T_h\ln\left({\kappa_1\over \kappa_2}\right)-T_c\ln \left(\frac{\kappa_1}{\kappa_2}\right)\right]}
{{k_{_B} \over 2}T_h \left[ \ln\left({\kappa_1\over \kappa_2}\right)+2(1-\rho_{_S})\left(1-\frac{T_c}{T_h}\right)\right]}=\frac{\left(1-\frac{T_c}{T_h}\right)\ln\left(\frac{\kappa_1}{\kappa_2}\right)}{\ln\left(\frac{\kappa_1}{\kappa_2}\right)+2(1-\rho_{_S})\left(1-\frac{T_c}{T_h}\right)}
\label{efiS} 
\ea
with $Q_{in}=Q_{AB}+(1-\rho_{_S})k_{_B}(T_h-T_c)$ where the parameter $\rho_{_S}$ considers a possible regeneration mechanism, when $\rho_{_S}=1$ there is a perfect regeneration and, $\rho_{_S}=0$, implies non-regeneration. As in the macroscopic model, when there is ideal  regeneration the Carnot efficiency is recovered, otherwise, $\eta<\eta_C$.


\subsection{Ericsson-like Cycle}
Another cycle that can be constructed using a pair of isotherms is one that includes a pair of isobaric processes, so-called as Ericson cycle. For the system that concerns us in this work, an Ericsson-like cycle can be implemented as indicated in the figure \ref{ericsson}. Two quasistatic isothermal processes, where $\kappa$ changes but $T$ remains constant, and two quasistatic isobaric processes where both $\kappa$ and $T$ change simultaneously. The four stages of the Ericsson cycle can be stated as follows:

\begin{figure}
    \centering
    \includegraphics[scale=1.1]{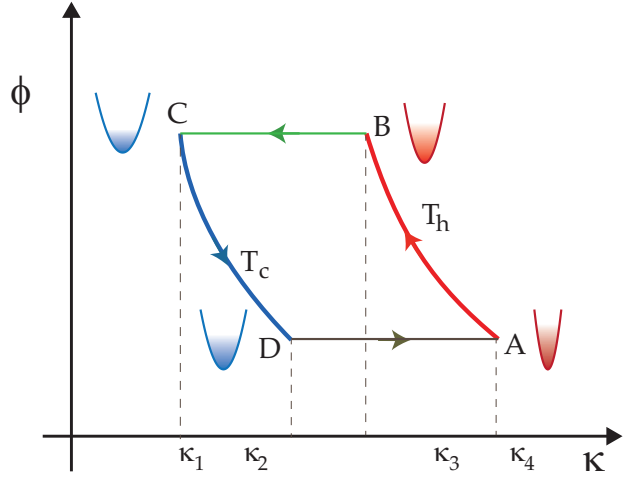}
    \caption{$\Phi$-$\kappa$ thermodynamic diagram of an Ericcson-like cycle where  (i) an isothermal expansion (red path); (ii) an isobaric expansion process (BC green path); (iii) an isothermal compression (blue path) and (iv) an isobaric compression process (DA gray path).}
    \label{ericsson}
\end{figure}

i) Isothermal expansion process ($A\to B$): The optical trap expansion-space is heated at $T_h={\rm const}$, changing from $\kappa_4\to \kappa_3$ ($\kappa_3< \kappa_4$). Besides, ($\Delta E)_{AB}=0$ and $Q_{AB}=-W_{AB}$, see Eqs. (\ref{wab}) and (\ref{cqab}).

ii) Isobaric expansion process ($B \to C$): The expanded space for a Brownian particle changes from $\kappa_3\to \kappa_1$, ($\kappa_1 <\kappa_3$) and picks up heat at a high constant-$\Phi_u$ value from $T_h \to T_c$. Then, from Eqs. (\ref{Wa}) and (\ref{Qa})
\ba
W_{BC} &=& \Phi_u\int _{\kappa_3}^{\kappa_1} {d\kappa} = 
 \Phi_u\left(\kappa_1- \kappa_3\right)=\frac{k_{_B}}{2}\left(T_c-T_h\right)<0 ,   \label{wabE} \\
 \nonumber\\
 Q_{BC}&=& {k_{_B}\over 2}\int_{T_h}^{T_c}{dT} = {k_{_B}\over 2}\left(T_c-T_h\right) <0. \label{cqabE}
 \ea
where the equation-like state $2\Phi =\langle x^2 \rangle$, has been used to obtain Eq. (\ref{cqabE}). 
 
iii) Isothermal compression process ($C \to D$): The compression space for the colloidal particle is cooled at $T_c={\rm const}$, modifying $\kappa$ from $\kappa_1\to \kappa_2$ ($\kappa_2 > \kappa_1$), as well as $Q_{CD}=-W_{CD}$, see Eqs. (\ref{cwcd}) and (\ref{cqab1}).

iv) Isobaric compression process $D\to A$): The compressed space of the optical trap $\kappa_2\to \kappa_4$  ($\kappa_4 > \kappa_2$) is carried out by decreasing the thermal environment at low constant- $\Phi_l$ value from $T_c \to T_h$. Thus,
\ba
W_{DA} &=& {\Phi_l}\int _{\kappa_2}^{\kappa_4} {d\kappa} = 
 {\Phi_l}\left(\kappa_4- \kappa_2\right)=\frac{k_{_B}}{2}\left(T_h-T_c\right)>0 ,   \label{wdaE} \\
 \nonumber\\
 Q_{DA}&=& {k_{_B}\over 2}\int_{T_c}^{T_
 h}{dT} = {k_{B}\over 2}\left(T_h-T_c\right)>0. \label{cqdaE}
 \ea
For this cycle, the total work performed is: $W_{tot} = W_{AB}+W_{BC}+W_{CD}+W_{DA}$. The total heat input to the cycle is $Q_{in}=Q_{AB}+  {1\over 2}k_{B}(T_h-T_c)$, the one absorbed in the hot isotherm and in the isobaric branch, in this case, as in the Stirling cycle, a possible heat regeneration mechanism can be proposed, in such a way that $Q_{in}=Q_{AB}+  {1\over 2}(1-\rho_{_E})k_{B}(T_h-T_c)$, where $\rho_{_E}$ is associated with the efficiency of the  regenerator  between isobaric branches. Thus, the expression for the efficiency for this Ericsson-like cycle reads:
 \begin{equation}
     \eta= \frac{W_{tot}}{Q_{in}}=\frac{\frac{k_B}{2}\left[T_h \ln{\left(\frac{\kappa_1}{\kappa_2}\right)}-T_c \ln{\left(\frac{\kappa_4}{\kappa_3}\right)}\right]}{\frac{k_B}{2}T_h\left[ \ln{\left(\frac{\kappa_1}{\kappa_2}\right)}+(1-\rho_{_E})\left(1-\frac{T_c}{T_h}\right)\right]}=\frac{\left(1-{T_c\over T_h}\right)\ln \left(\frac{\kappa_1}{\kappa_2}\right)}{\ln \left(\frac{\kappa_1}{\kappa_2}\right)+(1-\rho_{_E})\left(1-{T_c\over T_h}\right)},\label{etaEC}
 \end{equation}
since $\kappa_1=(T_c^2/T_h^2)\kappa_3$ and $\kappa_2=(T_c^2/T_h^2)\kappa_4$. Additionally, as in the Stirling cycle, $0\leq \rho_{_E} \leq 1$,  where $\rho_{_E} =1$ would emulate an ideal regeneration, while $\rho_{_E}=0$ implies null regeneration at the isobaric processes.

\section{low-dissipation methodology}

Once we have obtained from Eq. (\ref{dex2}) the equilibrium properties of the above three cycles,  we proceed to obtain the efficiency at maximum power for each out of equilibrium cycle, in which dissipative processes inevitably appear.
The newly state-of-the-art for optical trapping techniques use the pressure radiation and a focused beam of light to hold in place, or move small-scale objects inside low-density thermal environments \cite{Ashkin87,Ashkin187}. Despite the efforts through standard ways to control the temperature of environments \cite{Svoboda92,Leake04}, the kinetic contribution for the transport of colloids typically presents drift-diffusion processes, whose nature is purely dissipative. For a heat engine, if $\tau$ is the cycle duration, the average of extracted work is proposed as $\langle W\rangle= \langle W_{\infty}\rangle +\langle W_{dis}\rangle$, where $\langle W_{\infty}\rangle$ is the mean quasistatic work for longer cycle times and $\langle W_{dis}\rangle$ is the mean irreversibly dissipated work per cycle.

$\langle W \rangle=\langle W_{\infty}\rangle+\langle W_{irr}\rangle$ and the heat exchanged $\langle Q \rangle$ take into account the term $(\nicefrac{T\Sigma}{\tau}) > 0$, related to the ratio of entropy production (positive dissipation) per cycle. That is,
\begin{equation}
    \langle Q \rangle =\langle Q_{\infty}\rangle -\frac{T\Sigma}{\tau}, \label{avheat}
\end{equation}
where the subscript $\infty$ represents the average observable quantities at the quasistatic limit. In an analogous way to the irreversible phenomena that unavoidably takes place in macroscopic heat engines  operating at finite cycle times, the parameter $\Sigma$ represents the internal irreversibilites due to the coupling between the optical trapped particle and the thermal environment. 
The out of equilibrium process  characterized by the macroscopic Eq. (\ref{dex2b}) with time-dependent stiffness $\kappa(t)$ and temperature $T(t)$ can be quantified by a dissipation factor $\Sigma$.
This way to describe the performance of energy conversion processes matches the low dissipation scheme for heat engines \cite{Esposito10,Gonzalez17}.

\subsection{Performance of an irreversible Carnot-like cycle}

In the model of Carnot cycle, the heat exchanges with the thermal baths only take place in the isothermal branches.
In this work, following the idea of Esposito et al. \cite{Esposito10}, an ensemble of colloidal particles in contact with thermal baths is now considered. But each particle is in contact with the reservoirs for a time $\tau_1$ at the hot branch and $\tau_2$ at the cold branch, with $\tau_1$ and $\tau_2$ being finite times. Under this scheme, there is an entropy production per cycle equals to
\begin{equation}
    \dot{\Sigma}_{N}=\frac{\Sigma_1}{\tau_1}+\frac{\Sigma_2}{\tau_2} \label{totentrop}
\end{equation}
The quasistatic regime is reached when $\tau_1\to \infty$ as well as $\tau_2 \to \infty$. That is, there is total exchanged heat due to an amount of dissipated energy for each process. Thus,
\begin{equation}\label{qsnet}
\begin{split}
    Q_1 & =T_c\left(-\Delta S-\frac{\Sigma_1}{\tau_1}\right) \\
    Q_2 & =T_h\left(\Delta S-\frac{\Sigma_2}{\tau_2}\right).
\end{split}
\end{equation}
In the weak dissipation approximation  $\Sigma_1$ and $\Sigma_2$ express the increase of dissipated energy when the processes are carried out at finite time. Then, from eqs. (\ref{cqab})  and (\ref{cqab1})  the power output for this Brownian Carnot-like cycle is given by
\begin{equation}
    P\equiv \frac{-W}{\tau}=\frac{(T_h-T_c)\Delta S-\left(\frac{T_c\Sigma_1}{\tau_1}+\frac{T_h\Sigma_2}{\tau_2}\right)}{\tau_1+\tau_2}=\frac{\frac{k_B}{2}(T_h-T_c)\ln{\left(\frac{\kappa_1}{\kappa_2}\right)}-\left(\frac{T_c\Sigma_1}{\tau_1}+\frac{T_h\Sigma_2}{\tau_2}\right)}{\tau_1+\tau_2}. \label{potcarn}
\end{equation}
Under these assumptions, a system can achieve the so-called maximum power output regime, when the derivatives of $P$ with respect to $\tau_1$ and $\tau_2$ equal to zero. After substituting Eqs. (\ref{cqab}) and (\ref{cqab1}) into Eq. (\ref{potcarn}), the physical attainable solution for $\tau_1$ and $\tau_2$ are
\begin{equation}\label{tsphys}
\begin{split}
    \tau_1^{*} & =\frac{2T_c \Sigma_1}{(T_h-T_c)\Delta S}\left(1+\sqrt{\frac{T_h \Sigma_2}{T_c \Sigma_1}}\right) \\
    \tau_2^{*} & = \frac{2T_h \Sigma_2}{(T_h-T_c)\Delta S}\left(1+\sqrt{\frac{T_c \Sigma_1}{T_h \Sigma_2}}\right).
\end{split}
\end{equation}
The same expressions founded by Esposito et al. \cite{Esposito10} for a traditional heat engine performing finite-time Carnot cycles.  By considering Eqs. (\ref{qsnet}) and (\ref{tsphys}), as well as the expression for the efficiency [see Eq. (\ref{cea})], the efficiency at maximum power regime is obtained as follows:
\begin{equation}
    \eta_{_{MP}}=\frac{(T_h-T_c)\left(1+\sqrt{\frac{T_c \Sigma_1}{T_h \Sigma_2}}\right)}{T_h\left(1+\sqrt{\frac{T_c \Sigma_1}{T_h \Sigma_2}}\right)^2+T_c\left(1-\frac{\Sigma_1}{\Sigma_2}\right)}. \label{effMP}
\end{equation}
For a symmetric dissipation case $(\Sigma_1=\Sigma_2)$, the well-known Curzon-Ahlborn efficiency is recovered.
\begin{equation}
    \eta_{_{MP}}=1-\sqrt{1-\eta_{_C}}=1-\sqrt{T_c\over T_h} \equiv\eta_{_{CA}}.\label{etaCA}
\end{equation}
\begin{figure}
    \centering
    \includegraphics[scale=0.80]{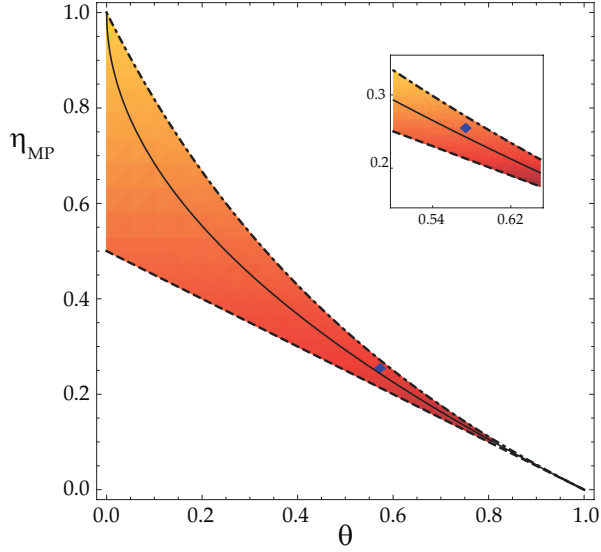}
    \caption{Energetic performance map at maximum power of an irreversible Carnot-like cycle as function of $\theta$. The well-known Curzon-Ahlborn efficiency is denoted by the solid line. The upper and lower bound for the asymmetric cases are marked by a dot-dashed line and a dashed line, respectively.}
    \label{fig:mapetCC}
\end{figure}
 On the other hand, when asymptotic asymmetric cases are considered, $\nicefrac{\Sigma_1}{\Sigma_2}\to 0$, lead $\eta_{MP}$ tends to an upper bound, namely $\eta_{_{MP}}^{u}=\nicefrac{\eta_C}{2-\eta_C}$. Likewise, if $\nicefrac{\Sigma_1}{\Sigma_2}\to \infty$ then $\eta_{MP}$ tends to a lower bound: $\eta_{MP}^{l}=\nicefrac{\eta_C}{2}$. Fig. \ref{fig:mapetCC} shows $\eta_{MP}$ as function of $\theta$ (where $\theta=T_c/T_h$), and the physically attainable region for the performance of these Brownian systems within the weak dissipation approach in the same way as in \cite{Chen89,Gaveau10}. The experimental value ($\theta \approx 0.57$, $\eta_{MP} \approx 0.25$) obtained for the efficiency  at maximum power of the Carnot cycle developed with a Brownian particle \cite{Martinez16} is also shown, and  it is located in the physical attainable region  (see Fig. [\ref{fig:mapetCC}]). This shows that considering a low dissipation model, where all irreversibilities are quantified through the isothermal branches and the adibatic branches as instantaneous, is consistent with the experimental results.

\subsection{Performance of irreversible Stirling and Ericsson-like cycles}

The other two symmetric block-thermodynamic cycles (Stirling and Ericsson cycles), can be also studied within the weak dissipation approximation.  From \cite{Blickle12}, the experiment for a stochastic Stirling heat engine, there is reported that it had been performed along two isothermal processes and,  due to the difficulties to keep hot and cold reservoirs thermally isolated the temperature of the surrounding liquid is suddenly changed in the isochoric processes. Therefore, considering the sources of irreversibilities only in the isothermal branches is not an unviable assumption and the approach of a low dissipation model can be considered for this cycle. Similarly, for the Ericcson cycle it will be assumed that the irreversibilities come from the isothermal branches, while in the isobaric type the thermal bath and the potential stiffness are changed simultaneously and instantaneously. On the other hand, the total ratio of entropy production per cycle of the protocol evolution is expressed by Eq.(\ref{totentrop}).

 Due to the symmetry along isochoric and isobaric processes, the same expression for the entropy transfer as in the Brownian-Carnot cycle is recovered, and therefore the power output for these Brownian Stirling-like and Ericsson-like cycle model [see Eq. (\ref{potcarn})].

For the case of a Brownian Stirling cycle and after substituting Eqs. (\ref{qsnet}) and (\ref{tsphys}) into Eq.(\ref{efiS}), the efficiency at maximum power output reads:
\begin{equation}
    \eta_{_{MP}}^{S} = \frac{\left(T_h-T_c\right) \left( 1 +\sqrt{\frac{T_h\Sigma_2}{T_c\Sigma_1}}\right)}{T_h\left(1+\sqrt{\frac{T_c \Sigma_1}{T_h \Sigma_2}}\right)^2+T_c\left(1-\frac{\Sigma_1}{\Sigma_2}\right) +4 (1-R_{_S})(T_h-T_c)\left( 1 + \sqrt{\frac{Tc\Sigma_1}{T_h\Sigma_2} } \right)},\label{etmaxpS}
\end{equation}
where $(1-R_{_S})=(1-\rho_{_S})\left[\ln{\left(\nicefrac{\kappa_1}{\kappa_2}\right)}\right]$. For the symmetric dissipation case $(\Sigma_1=\Sigma_2)$, the efficiency at maximum power output of a Stirling-like cycle is:
\begin{equation}
    \eta_{_{MP}}^{SS}=\frac{\eta_{_{CA}}}{1+4 (1-R_{_S})\eta_{_{CA}}}.\label{etmaxpSS}
\end{equation}

The asymmetric cases for this cycle represent two limits, the first $\left(\nicefrac{\Sigma_1}{\Sigma_2}\right)\to 0$, leads to the upper bound $\eta_{MP}^{+S}=\nicefrac{\eta_C}{\left[2(1+2 (1-R_{_S})\eta_C)-\eta_C\right]}$. On the other hand, when $\left(\nicefrac{\Sigma_1}{\Sigma_2}\right)\to \infty$, the lower bound is $\eta_{MP}^{-S}=\nicefrac{\eta_C}{2\left[1+2 (1-R_{_S})\eta_C\right]}$.

\begin{figure}
    \centering
    \includegraphics[scale=1.0]{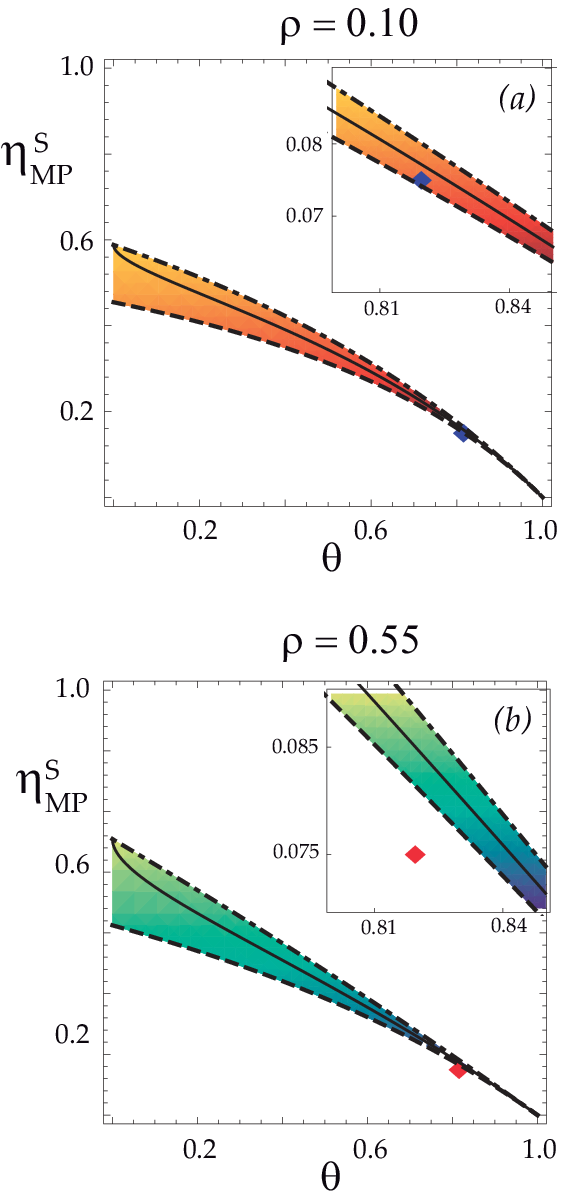}
    \includegraphics[scale=1.0]{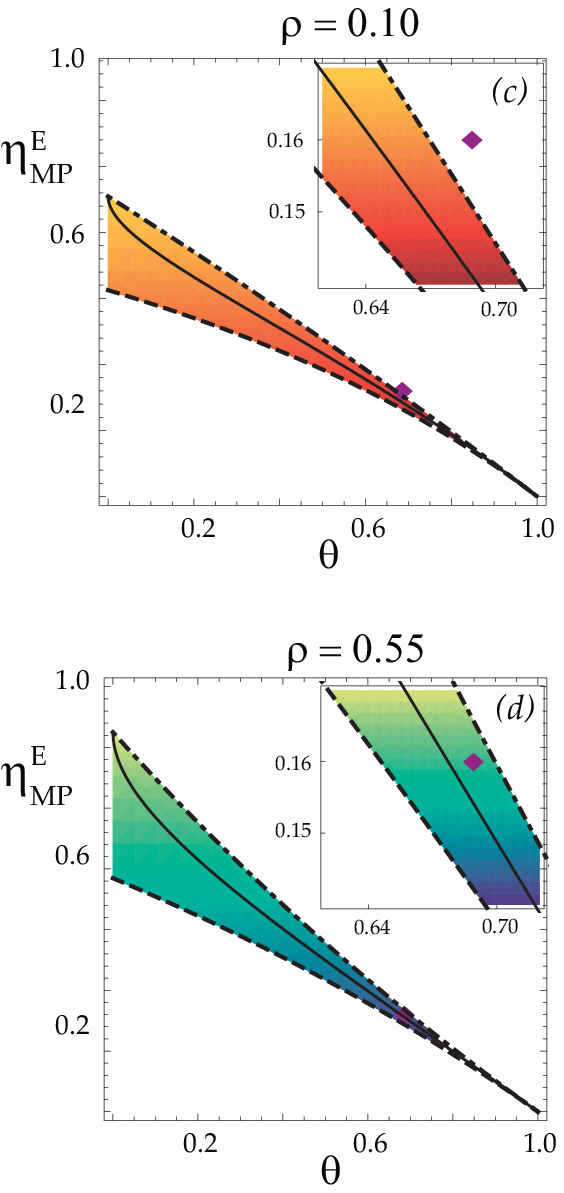} \par
    \caption{Energetic performance map at maximum power of two irreversible cycles (Stirling-like and Ericsson-like ones) as a function of $\theta$. The graphs in (a) and (c) emulate  arbitrary low regeneration-like processes ($\rho_{_S}=\rho_{_E}=0.1$), while (b) and (d) represent arbitrary high regeneration ones ($\rho_{_S}=\rho_{_E}=0.55$). The efficiency for symmetric dissipation cases is denoted by the solid line. Likewise, the upper and lower bound for the asymmetric cases are marked by a dot-dashed line and a dashed line, respectively.}
    \label{fig:mapetEC}
\end{figure}

Figs. \ref{fig:mapetEC}.a and \ref{fig:mapetEC}.b depict the behavior of $\eta_{MP}$ as function of $\theta$. The energetic performance of these Brownian systems is sketched for two cases: In a) it is emulated a low regeneration process ($\rho_{_S}=0.1$) while in b) a high regeneration one ($\rho_{_S}=0.55$). Now, the experimental value ($\theta \approx 0.82$, $\eta_{MP}^{S} \approx 0.075$; denoted by blue and red diamonds) for a Stirling-like cycle was obtained by \cite{Blickle12}, and it is in agreement with the physical attainable region for a low regeneration process (Fig. \ref{fig:mapetEC}.a).

Likewise, the efficiency at maximum power output regime for a Ericsson-like cycle is obtained from Eqs.  (\ref{etaEC}), (\ref{qsnet}) and (\ref{tsphys}),
\begin{equation}
    \eta_{MP}^{E}=\frac{(T_2-T_1)\left(1+\sqrt{\frac{T_1 \Sigma_1}{T_2 \Sigma_2}}\right)}{T_2\left(1+\sqrt{\frac{T_1 \Sigma_1}{T_2 \Sigma_2}}\right)^2+T_1\left(1-\frac{\Sigma_1}{\Sigma_2}\right) +2(1-R_{_E})(T_2-T_1)\left( 1 + \sqrt{\frac{T1\Sigma_1}{T_2\Sigma_2} } \right)},\label{etmaxpE}
\end{equation}
where $ (1-R_{_E})=(1-\rho_{_E})\left[\ln{(\nicefrac{\kappa_1}{\kappa_2})}\right]^{-1}$. By assuming the symmetric dissipation $(\Sigma_1=\Sigma_2)$, the efficiency $\eta_{MP}^{E}$ is:
\begin{equation}
    \eta_{MP}^{ES}=\frac{\eta_{CA}}{1+2(1-R_{_E})\eta_{CA}}.\label{etmaxpES}
\end{equation}

Now, the asymmetric cases can also be studied, the first $\left(\nicefrac{\Sigma_1}{\Sigma_2}\right)\to 0$, leads to the upper bound $\eta_{MP}^{+E}=\nicefrac{\eta_C}{\left[2(1+ (1-R_{_E})\eta_C)-\eta_C\right]}$. On the other hand, when $\left(\nicefrac{\Sigma_1}{\Sigma_2}\right)\to \infty$, the lower bound is $\eta_{MP}^{-E}=\nicefrac{\eta_C}{2\left[1+ (1-R_{_E})\eta_C\right]}$.

 In the same way, Figs. \ref{fig:mapetEC}.c and \ref{fig:mapetEC}.d also depict the behavior of $\eta_{MP}$ as function of $\theta$. The energetic performance for a proposal of Brownian systems with a Ericsson-like cycle can be sketched for the same two cases: In c), a low regeneration process ($\rho_{_E}=0.1$), in d) for a high regeneration one ($\rho_{_E}=0.55$). Note, the general expressions for the performance of Brownian engines at maximum power (see Eqs.(\ref{effMP}), (\ref{etmaxpS}) and (\ref{etmaxpE})) show an energetic hierarchy namely, $\eta_{MP}^{S}<\eta_{MP}^{E}<\eta_{MP}$. The value ($\theta \approx 0.69$, $\eta_{MP}^{E} \approx 0.16$; denoted in purple diamonds) shown in Figs. \ref{fig:mapetEC}.c and \ref{fig:mapetEC}.d has been inferred through a linear interpolation; that is, as $\eta=\eta(\theta)$, it is possible to determine the equation of the line passing through $(\theta_{MP}^S,\,\eta_{MP}^S)$ and $(\theta_{MP},\,\eta_{MP})$; therefore, a qualitative $(\theta_{MP}^E,\,\eta_{MP}^E)$ can be obtained. Then, by considering the same Brownian system in [\cite{Martinez16, Blickle12}], it can be determined the guidelines to emulate an Ericsson-like cycle under analogous thermodynamics conditions to the Carnot and Stirling-like ones.

Thus, Carnot, Stirling and Ericsson-like cycles exhibit different constrained performance maps (see Fig. \ref{fig:mapetEC}). These graphs explicitly show that the contribution of the absorbed heat by the isochoric and isobaric paths limit the performance of the Brownian system. Moreover, the existence of a controlled regeneration system would make it possible to compare the maximum power developed by both cycles with the Carnot one. \\

\section{Conclusions} 
 In this paper, the strategy of low dissipation approach for macroscopic heat engines \cite{Esposito10}, has been used to calculated the efficiency at maximum power of three Brownian heat engines (Carnot, Stirling, and Ericsson-like cycles). In our case, the irreversible work and heat averages exchanged by the system with the heat bath are proposed to satisfy that  $\langle W\rangle \approx \langle W\rangle_{\infty}+ \frac{\Sigma}{\tau}$ and  $\langle Q\rangle\approx \langle Q\rangle_{\infty}-\frac{T\Sigma}{\tau}$, where $\langle W\rangle_{\infty}$ and 
$\langle Q\rangle_{\infty}$ are  the work and heat averages under equilibrium conditions, respectively. The equilibrium quantities are calculated through a state-like equation associated with $\langle x^2\rangle$, coming from a macroscopic Eq. (\ref{dex2b}) and  this latter from  Langevin Eq. (\ref{odho}). In our proposal we do not use a specific protocol for $\kappa(t)$ and $T(t)$ to quantify the non-equilibrium thermodynamic quantities, instead, we take advantage of low dissipation approach and use the dissipation parameter $\Sigma$ to calculate the efficiency at maximum power for the three aforementioned Brownian heat engines. Inspired by the experiment carried out by Blickle and Bechinguer \cite{Blickle12}, we consider that the pair of non-isothermal branches in each cycle are instantaneous. That is, for the Carnot-like cycle the adiabatic ones, for the Stirling-like cycle the isochorics and for the Ericsson-like cycle the isobarics. And therefore the irreversible effects are only taken into account in the two isothermal processes.

Our proposal shows that in the case of a stochastic Carnot-like heat engine, the efficiency at maximum power is quite similar to the one
reported in \cite{Esposito10} for macroscopic heat engines. Also, 
in the case for which $\Sigma_1 = \Sigma_2$, it recovers the well known Curzon-Alborn efficiency. While the limits for the asymmetric case an efficiency region is delimited (see Figs. \ref{fig:mapetCC} and \ref{fig:mapetEC}), which in the cases of Stirling and Ericsson-like cycles depend on the possible so-called regeneration mechanism (Fig. \ref{fig:mapetEC}). 
From the experimental works that have been carried out for the Carnot and Stirling-like cycles \cite{Martinez16,Blickle12}, interpolates a possible experimental value that would reproduce an Ericsson-like cycle within the interval $0.57 \leq \theta \leq 0.82$. Additionally, a hierarchy for the efficiencies of the three cycles were found ($\eta_{MP}^{S}<\eta_{MP}^{E}<\eta_{MP}$).

\begin{acknowledgments}

OCV thanks the CONACyT-M\'exico scholarship. NSS thanks to SIP-IPN (M\'exico).

\end{acknowledgments}

\bibliography{biblio}
\vskip3.0cm

\end{document}